\documentclass[11pt,twoside]{article}

\usepackage{asp2006}
\usepackage{epsf}
\usepackage{lscape}

\usepackage{graphicx}

\def\kms {\hbox{km{\hskip0.1em}s$^{-1}$}} 
\def\ee #1 {\times 10^{#1}}          
\def\ut #1 #2 { \, \mathrm{#1}^{#2}} 
\def\u #1 { \, \mathrm{#1}}          
\def\msol{\hbox{$\hbox{M}_\odot$}}

\def\kms{km s$^{-1}$}

\begin{document}
\title{Massive Star Formation Near Sgr A* and Bimodal Star Formation in the Nuclear Disk}   
\author{F. Yusef-Zadeh$^1$ \& M.Wardle$^2$}   
\affil{$^1$Department of Physics and Astronomy, Northwestern University, Evanston,
IL 60208  USA \break email: zadeh@northwestern.edu\\
$^2$Department of Physics, Macquarie University, Sydney NSW 2109,
Australia  \break email: wardle@physics.mq.edu.au\\}


\begin{abstract}

The history of star formation in the strong gravitational potential of
the Galactic center  has been of much interest, recently. 
We propose that the sub-parsec-scale disk of massive stars orbiting the
massive black hole at the Galactic center can be interpreted in terms
of partial accretion of extended Galactic center clouds, such as the
50 \kms molecular cloud, as these clouds envelop Sgr A* on their
passage through the inner Galactic center.  The loss of angular
momentum of the captured cloud material by self-interaction subsequent
to gravitationally focusing by Sgr A* naturally creates a compact
gaseous disk of material close to Sgr A* in which star formation takes
place.  On a larger scale the formation of massive clusters such as
the Arches and Quintuplet clusters or on-going massive star formation
such as Sgr B2 could also be triggered by cloud-cloud collisions due
to gravitational focusing in the deep potential of the central bulge.

Unlike the violent and high-pressure environment of clustered star
formation triggered by cloud-cloud collision, there are also isolated
pockets of star formation and quiescent dense clouds.  These sites
suggest an inefficient, slow mode of star formation.  We propose
enhanced cosmic rays in the nuclear disk may be responsible for
inhibiting the process of star formation in this region.  In
particular, we argue that the enhanced ionization rate due to the
impact of cosmic-ray particles is responsible for lowering the
efficiency of on-going star formation in the nuclear disk of our
Galaxy.  The higher ionization fraction and higher thermal energy due
to the impact of these electrons may also reduce MHD wave damping
which contributes to the persistence of the high velocity dispersion
of the molecular gas in the nuclear disk.

\end{abstract}


\section{Introduction}   

Radio and infrared observations of the central region of our Galaxy
have come a long way since two major discoveries were reported more
than 20 years ago, namely, the central stellar cluster at the Galactic
center (Becklin \& Neugebauer 1968) and the compact radio source Sgr
A* (Brown and Balick 1974).  After all these years, one wonders who
would have predicted that the stars of the young central cluster at
the Galactic center are on Keplerian orbits about Sgr A*.  We now know
from near-IR measurements of Sgr A* combined with proper motion
measurements of Sgr A* (Reid \& Brunthaler 2004; Ghez et al.  2005;
Eisenhauer et al.  2005) that a massive black hole of 3--4$\times10^6$
\msol is coincident with Sgr A*.  Again, who would have realized that
most of the cluster members are distributed in one or perhaps two
stellar disks orbiting Sgr A* (Paumard et al.  2006; Lu et al.  2006)?
Again, who would have thought that in the hostile environment of the
nucleus of our Galaxy,  two additional young clusters (i.e. the
Arches and Quintuplet clusters) lie within a projected distance of 30 pcs
from Sgr A* (e.g., Cotera et al.  1996; Figer et al.  1999).  Lastly,
on a larger scale, who would have predicted the co-existence of two
star forming sites with extreme differences in star formation rates;
namely Sgr B2 and a ridge of infrared dark clouds with quiescent star
formation (Lis \& Carlstrom 1994).

There are many challenges to understanding the star formation history
in the nucleus of our Galaxy.  Here we attempt to elucidate the
environmental factors affecting star formation in the complex and rich
region of our Galactic center.  We argue that enhanced cloud-cloud
collisions and cosmic rays, are likely to be important in on-going and
past star formation activity in this region.  In particular, we
discuss two modes of star formation in the nuclear disk of our Galaxy.
One mode is induced by cloud-cloud collision in the nuclear disk which
we claim to be responsible for cluster star formation.  The other mode
is analogous to that of low-mass star formation in which
gravitationally unstable cores within a cloud contract to the point
that they overwhelm their magnetic support and form isolated stars.

\section{Cluster Star Formation } 

\subsection{The Central Cluster near Sgr A*}

Proper motion and spectral line studies of stars within 0.5 pc of Sgr
A* have recently shown that a cluster of about 80 massive stars in the
inner pc of the Galaxy reside in one, and possibly a second, rotating
disk (Paumard et al.  2006; Lu et al.  2006).  These disks, inferred
to have masses about 10$^4$ \msol, have well-defined inner edges and
are counter-rotating at large angles from each other with typical
stellar ages of 6$\pm2$ million years (Paumard et al.  2006).  They
are somewhat disordered with h/r$\sim$0.1 and stellar orbit
eccentricities ranging roughly between 0.2 and 0.8.  Two scenarios
have been proposed to explain the origin of these young stars, in-situ
star formation within a disk (Nayakshin et al.  2007), or dynamical
migration of stars formed at a large distance (Gerhard 2001).  The
latter requires a high mass concentration for dynamical friction to be
effective on a time scale of a few million years (Hansen \&
Milosavljevi\'c\ 2003; Kim, Figer \& Morris 2004; G\"urkan \& Rasio
2005), perhaps provided by an IMBH associated with the initial cluster
of stars.  Alternatively, in the in-situ picture, the young stars are
formed by the fragmentation of gaseous disk that has been captured by
the strong gravitational potential of Sgr A* (Nayakshin 2006).
Nayakshin et al.  (2007) model the fragmentation of a gravitational
unstable gaseous disk to form massive stars.  This scenario needs to
address the issue of the observed high eccentricity of stars in a
relatively thick accretion disk, as it is difficult to develop high
eccentricity from an initially circular orbits in a thin accretion
disk.  Furthermore, this scenario leaves the question of how the
orbiting gaseous disk got there in the first place unanswered.
Capturing a cloud passing to one side of Sgr A* turns out to run into
difficulty because the cloud has no way of getting rid of its angular
momentum, requiring an unlikely situation in which a dense and very
compact cloud is on a trajectory towards Sgr A* with essentially zero
impact parameter (Wardle and Yusef-Zadeh 2007).

\begin{figure}
\centering
\includegraphics[scale=0.4,angle=0]{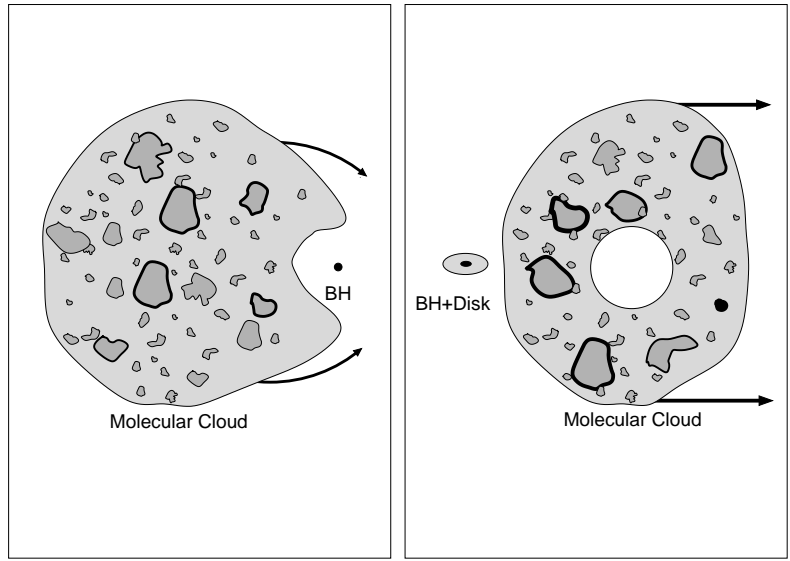}
  \caption{A schematic diagram of a cloud impacting the Sgr A*.  The
  panel to the left shows the effect of gravitational focusing in
  capturing a colliding gaseous material.  The panel to the right
  shows the carved out inner region of the cloud is captured first before 
 brought  in closer to  Sgr A*. The outer region of the cloud
continues its motion in the direction away from Sgr A*.}
\label{fig:contour}
\end{figure}  

These issues can be circumvented by considering a more common event:
the near-radial passage of an extended cloud that sweeps through the
strong gravitational potential of the Galactic center and temporarily
engulfs Sgr A*.  Sgr A* and the evolved stellar cluster with r$^{-2}$
density profile dominate the gravitational potential within and beyond
the inner pc, respectively.  In this picture, the gravitational
potential more strongly deflects the inner regions of the cloud,
affecting a collision between fluid elements that pass on either side
of Sgr A* and oppositely directed angular momenta.  This is a messy
version of Hoyle-Lyttleton accretion (Edgar 2004) because of the pronounced
inhomogeneity in the approaching molecular cloud.  The resulting
dissipation permits this gas to become bound to Sgr A*.  Furthermore,
the captured material is brought in from a capture radius of $\sim$3
to $\sim$0.3 pcs.  The loss of angular momentum forms a compact disk
and creates a large disk surface density which becomes gravitationally
unstable as the gas settles down and cools.  The details of the
inhomogeneities in the cloud and its initial trajectory determine the
direction of the disk's angular momentum which as a result is largely
unrelated to orientation of Galactic rotation.  Simple estimates
(Wardle \& Yusef-Zadeh 2007) show that the disk becomes gravitationally
unstable before settling down and cooling to the point of becoming
thin, thus the stars that are formed should have a range of
eccentricities.  Both the clockwise and counter-clockwise stellar
disks can also be explained in the context of this model by either the
inhomogeneous nature of a single giant molecular cloud or by the
passages of two different clouds separated by $\sim10^6$ years.  These
events may also coincide with the growth of the black hole as well as 
 with  an increase in the luminosity due to the
accretion of gas directly onto Sgr A*, two subjects that are  not discussed
here.

While the inner region of the cloud interacts with Sgr A*, the outer
regions continue their journey along the original radial orbit.
Figure 1 illustrates the shape of the cloud before and after the cloud
sweeps the Galactic center.  The well-known 50 \kms\ cloud at the
Galactic center  may be related to the formation of the current stellar
disk: there is strong evidence that it is interacting with the Sgr A
East supernova remnant which itself lies behind Sgr A*.  Sgr A East
and the 50 \kms\ cloud are thought to lie within roughly ten parsecs
of Sgr A* (e.g., Melia et al.  2001; Yusef-Zadeh et al.  2001).  If
this cloud extends for $\sim$25--50 pcs, it would take
$\sim 0.5-1\times 10^6$ years before a inner portion of the cloud is
completely carved out.  This time scale is consistent with that needed
to form one or two disks of stars.  A more detailed account of this
scenario can be found elsewhere (Wardle \& Yusef-Zadeh 2007).

\subsection{The Orbiting  Molecular Ring}

On a scale greater than 0.5 pc, we find a disordered molecular ring
that orbits Sgr A* at a distance of 2--10 pc with a velocity of 110
km/s (e.g. Genzel \& Townes 1987; Jackson et al.  1993).  This
circumnuclear molecular ring or disk (CND) surrounds a cavity of
ionized gas, known as Sgr A West.  The ring is quite messy,
incomplete, partially collisionally excited (e.g., Yusef-Zadeh et al.
2001) and is tilted with respect to the Galactic plane.  This ring of
gas may be a relic from a passage of a cloud similar to that
envisioned in the previous section (see the illustration in Fig.  1),
see also (Sanders 1998), but with, for example, a lower cloud speed resulting
in the capture of a larger region of the incoming cloud.  
The lack or the presence of star formation  can be understood in
the context of inhomogeneity of the radially moving cloud toward Sgr A*.

It is not clear whether there is any
signature of on-going massive star forming activity in the molecular
ring.  
The lack of on-going star
formation implies that the molecular surface
density   is not  sufficiently high  for the gravitational instability to take place.
On the other hand, recent high resolution molecular observations of the CND show
dense molecular gas with H$_2$ density 10$^7$ cm$^{-3}$ (Christopher
et al.  2005).  This highly clumped molecular material can withstand
the strong tidal effects of the Galactic center if n$_{H2} >
10^7\times (1.6pc/r)^{1.8}$ cm$^{-3}$ and could potentially collapse
and form a new generation of massive stars.  It is puzzling,
therefore, that no sites of on-going star formation have been
identified in the molecular ring. 

%

\subsection{The Arches and Quintuplet Clusters \& Sgr B2}

One of the question that we'd like to examine is whether
cloud-cloud collisions are relevant to the formation of other
clusters in the Galactic center.  The Arches and Quintuplet clusters
are located at a projected distance of about 30 pc.  It is not clear
exactly the location of the parent cloud from which these clusters are
formed from.  However, it is clear that these clusters are not formed
from a disk of gaseous material.  This is because the orbital time
scale is long at large distances from Sgr A* when compared to the
free-fall time scale.

Although the production of a disk of gas is not effective beyond the
inner several parsecs of Sgr A*, radially moving clouds can be focused
by the gravitational potential of the evolved stellar cluster and
enhance the cross-section for cloud-cloud collision.  There is
mounting evidence that the population of molecular clouds in the
nuclear disk show forbidden and high velocities.  This unusual
kinematics can be explained by the tidal torque of the barred
potential of the nuclear bulge or by the tidal friction of the bulge
stars on clouds (e.g., Bally et al.  1988; Stark et al.  1991; Morris
and Serabyn 1996).  These effects can lead to rapid inward motion of
clouds toward the Galactic center on highly elliptical orbits.  The
well-known -30 \kms\ cloud with its forbidden velocity on the positive
longitude side is thought to be associated with a star forming region
(e.g.\ Zhao et al.  1993).

The compression associated with violent collisions of clouds creates a
high pressure environment suitable for cluster star formation (Tan and
McKee 2002) provided that gravity is stronger than the tidal shear
associated with the Galactic center  potential.  This implies that
cluster star formation through this process will be inefficient within
a few tens of parsec of Sgr A*, but may be effective beyond that.
Indeed, there are several dense and massive ammonia clumps associated
with the 45 \kms molecular cloud in the Sgr A complex (e.g., Serabyn
\& G\"usten 1987) that show no signs of on-going stars formation.  On
the other hand, there is spectacular massive star formation with a
dense cluster of ultracompact HII regions associated with Sgr B2 at a
projected distance of $\sim75$ pcs.  In fact, previous analysis of
molecular clouds from several studied suggest that massive star
formation in Sgr B2 is triggered by the collision between the 65 and
80 \kms molecular clouds.  (Mehringer et al.  1993; Hasagawa et al.
1994).
	
\section{Isolated Star Formation } 

Most of the discussion in previous sections focused on bursts of
massive star formation in the Galactic center region.  However,
several cases of isolated massive star formation or of quiescent star
formation have been noted in the Galactic center region.  For example,
on a scale of five to ten parsecs from Sgr A*, there is a well-known
cluster of four compact HII regions that lie at the edge of the 50
km/s molecular cloud M-0.02-0.07.  These HII regions are the closest
known sites of on-going massive star formation in the Galactic center.
The HII regions are excited by O8-9 stars (e.g., Goss et al.  1985).
Mid-IR spectroscopic measurements of these HII regions have detected
[NeII] line emission from all four components (A-D) at radial
velocities around 40 km s$^{-1}$ (Serabyn, Lacy \& Achtermann 1992).
Another star forming site, SgrA-E, F, and G, within 5 pc of Sgr A* is
known to be associated with the 20 km s$^{-1}$ molecular cloud
M--0.13--0.08.  SgrA-E and F are known to be nonthermal features.
Interestingly,  all three sources SgrA-E--F and the bright circular HII
feature SgrA-G to the southwest corner coincide with the peak of
molecular line emission.  The HII feature is thought to be excited by
a massive star (Ho et al.  1985).  There is also a great deal of
quiescent star formation in a ridge of molecular clouds  (Lis \& Carlstrom 1994).  
These clouds show no signs of active star formation despite being characterized as
members of the population of Galactic center molecular clouds.  The
question that we'd like to raise is whether there exists another star
formation mechanism that can be applied to these sites.  Unlike the
efficient mode of star formation induced by cloud-cloud collisions,
this mechanism which is the upscaled version of low-mass star formation in the 
disk (Shu, Adams \& Lizano 1987) has to be slow and inefficient in 
generating quiescent and isolated massive star formation.

\subsection{The Role of Cosmic Rays in Star Formation}

The interstellar medium of the central kpc (the nuclear disk) is
characterized by a strong concentration of molecular gas in the
nuclear disk or the so-called ``Central Molecular Zone'' (CMZ) (Morris
\& Serabyn 1996).  Physical conditions in this region are extreme,
characterized by high velocity dispersion ($\sim$20 km/s), high
density ($\sim10^4$ cm$^{-3}$) and high temperature molecular gas
($\sim70$K) (H\"uttemeister et al.  1993).  These clouds must have
high density in order to be gravitationally bound against the strong
gravitational shear that they experience in the gravitational
potential associated with the high stellar density in the central
region of the Galaxy.  The nature of on-going massive star formation
in the nuclear region is puzzling.

What could be the cause of the non-uniform star formation rate in the
Galactic center region?  One possibility is a spatially variable
enhanced flux of cosmic ray particles (Yusef-Zadeh, Wardle and Roy
2007).  Recent radio, infrared and X-ray measurements all point to an
enhanced cosmic-ray flux in the central region of the Galaxy.  In one
study, Oka et al.  (2005) reported strong H$_3^+$ absorption lines
toward several directions in the Galactic center region.  They infer
an unusually high column density of H$_3^+$ which implies that the
ionization rate in the nuclear environment must be more than two
orders of magnitude higher than that in the Galactic disk.  Also, a
study of H$_3$O$^+$ inferred an order of magnitude higher cosmic-ray
ionization rate toward Sgr B2 than in the disk (van der Tak et al.
2006).  In another study, the detection of low frequency 74 MHz radio
emission from the central disk of the Galaxy indicates that the
cosmic-ray electron density of the central 1.5$\times0.5$ degrees is
$\sim$7 eV cm$^{-3}$, about six times higher than that estimated
toward the inner 6$\times2$ degrees (LaRosa et al.\ 2005).  Lastly,
the fluorescent 6.4 keV K$\alpha$ line emission
throughout the Galactic center region can be accounted for by
the impact of low-energy cosmic-ray particles with neutral gas
(Yusef-Zadeh et al.  2007a).

The molecular gas temperature could be elevated by an enhanced
cosmic-ray flux in the nuclear disk (e.g. G\"usten \& Downes   1981) with
implications for star formation in this region.  The higher cloud
temperatures increase the Jeans mass, potentially changing the IMF in
a high pressure environment.  In addition a high cosmic-ray ionization rate
increases magnetic coupling to the cloud material, suppressing ambipolar
diffusion and increasing the time taken for gravitationally unstable
cores to contract to the point that they overwhelm their magnetic
support.  This implies that star formation is slowed down in the
nuclear region (Yusef-Zadeh, Wardle \& Roy 2007).  Although a low star
formation rate and an enhanced cosmic ray flux may contradict each other, we
believe the enhanced nonthermal particles arising from processes other
than SN shock acceleration.  For example, the production of excess
nonthermal particles may be related to the origin of nonthermal radio
filaments in the Galactic center (Nord et al.  2004; Yusef-Zadeh,
Hewitt \& Cotton 2004).

One additional consequence of an increased ionization fraction is that
waves are damped less strongly.  In a weakly ionized medium, waves
with frequencies $\omega \sim kv_A$ below the collision frequency of
neutral particles with ions, $\nu_{ni} = n_i <\sigma v>$, are damped
on a time scale $2 \nu_{ni} / \omega^2 $ (Kulsrud \& Pearce 1969;
Zweibel \& Josafatsson 1983), directly proportional to $n_i$.  The
power input required to maintain wave motions on a given scale is
reduced by the same factor.  If the damping of MHD waves is reduced,
then it may help to explain the observed high velocity dispersion of molecular
clouds observed in the nuclear disk (e.g. Bally et al.  1988; Martin
et al.  2004).

Another implication of the excess cosmic-ray particles is that
cosmic-ray heating of molecular gas can be achieved without raising
the dust temperature (G\"usten et al.  1981).  This could explain why
warm gas is not often accompanied by hot dust.  In the last two
decades, studies of this region indicate that the molecular gas is
warm, ranging between 75 to 200K (e.g., H\"uttemeister et al.  1993).
However, far-IR and sub-millimeter studies have shown a dust
temperature ranging between $\sim$13 to 40\,K (Odenwald \& Fazio 1984;
Cox \& Laureijs 1989; Pierce-Price et al.  2000).  In typical clouds
elsewhere in the disk of the Galaxy, the gas is heated by collisions
with warm grains that are heated by massive stars.  Thus, regions of
high kinetic temperature in star forming regions are strongly
correlated with clouds with high dust temperature.

A global heating mechanism such as cloud-cloud collisions has also
been suggested to explain the significantly higher gas temperature in
a large fraction of clouds in the nuclear disk (H\"uttemeister et al.
1993).  In fact there is evidence of shocked gas traced by the
detection of SiO emission from Galactic center molecular clouds
(Martin-Pintado et al. 1997) Although this is consistent with the
cloud-cloud collision picture proposed here, shocked emission cannot
heat molecular cloud in its entirety, because much of the shocked
emission is distributed in a thin layer where clouds collide with each
other.  In contrast, low energy cosmic rays can penetrate deep into a
cloud and heat the gas (Yusef-Zadeh, Wardle \& Roy 2007).

In summary, in order to explain the paradoxical nature of highly
efficiency cluster star formation co-existing with quiescent star
formation, we have proposed a bimodal distribution of star formation
in the nuclear region of our Galaxy.  Two environmental factors become
important in star formation processes in the nucleus of our Galaxy when compared to
those  of the disk.  One is the enhanced rate of cloud-cloud collision
due to the strong gravitational potential of the barred nuclear
potential. This colliding picture of clouds very close 
to the peak of the potential is a special case which can 
explain the origin of stellar disks  
orbiting Sgr A*.  The other is the 
enhanced cosmic rays 
permeated 
throughout
this region.  We believe these two factors play important roles in
enhancing as well as suppressing massive star formation in this region
of the Galaxy.





\begin{thebibliography}{}

\bibitem[]{}
Balick, B. \& Brown, R. L., 
1974, ApJ, 194, 265

\bibitem[]{}
Bally, J., Stark, A.A., Wilson, R.W. \& Henkel, C. 1988, ApJ,
324, 223

\bibitem[]{}
Becklin, E. E.  \& Neugenbauer, G. 1968, ApJ, 151, 145
     
     
\bibitem[Cox \& Laureijs(1989)]{1989IAUS..136..121C} Cox, P., \& Laureijs,
R.\ 1989, The Center of the Galaxy, 136, 121

\bibitem[]{}
Cotera, A. S. et al. 1996, ApJ, 461, 750

\bibitem[]{}
Christopher et al. 2005, ApJ, 622, 346

\bibitem[]{}
Edgar, R. 2004, New Ast Rev, 48, 843

\bibitem[]{}
Eisenhauer, F. et al.  2005, ApJ, 628, 246
     
\bibitem[]{}
Figer et al. 1999, ApJ, 514, 202

\bibitem[]{}
Genzel, R. \& Townes, C. H., 1987, ARA\&A, 25, 377

\bibitem[]{}
Gerhard, O. 2001, ApJL, 546, L39
     
\bibitem[]{}
Ghez, A. M. et al. 2005, ApJ, 620, 744

\bibitem[]{}
Goss et al. 1985, MNRAS, 215, 69p

\bibitem[]{}
G\"u¼rkan, M. A.  \& Rasio, F.  A. 2005, ApJ, 628

\bibitem[G\"usten \& Downes(1981)]
{1981A&A....99...27G} G\"usten, R. \&
Downes, D.\ 1981, \aap, 99, 27

\bibitem[]{}
Hansen, B. M. S. \&  Milosavljevi\'c‡,  S. ¡ApJ, 593, L77

\bibitem[]{}
Hasegawa, T., Sato, F., Whiteoak, J. B. \& Miyawaki, R. 1994, ApJ, 429, L77 

\bibitem[]{}
Ho et al. 1985, ApJ 288, 575

     
\bibitem[]{}
     {H\"uttemeister, S. et al.} 1993,
     {A\&A} 280, 255

\bibitem[]{}
Jackson et al. 1993, ApJ, 402, 173

\bibitem[]{}
Kim, S. S.,  Figer, D. F. \&  Morris, M. 2004, ApJ, 607, L123

\bibitem[]{}
Kulsrud, R. \& Pearce, W. P. 1969,
ApJ 156 445

\bibitem[]{}
     {LaRosa, T. N. et al.} 2005,
     {ApJ} 626, L23

\bibitem[]{}
     {Lis, D. C.\& Carlstrom, J. E.} 1994,
     {ApJ} 424, 189

\bibitem[]{}
     {Lis, D. C. \& Menten, K.} 1998,
     {ApJ} 507, 794

\bibitem[]{} {Lu, J. et al. 2006}, JPhCS, 54,  79

\bibitem[]{}
Martin-Pintado, J., de Vicete, P.,
Fuente, A. \& Planesas, P.  1997, ApJ, 482, L45

\bibitem[]{}
Mehringer, D. M., Palmer, P., Goss, W. M. \& Yusef-Zadeh, F. 1993, ApJ, 412, 684

\bibitem[]{}
Melia, F., Yusef-Zadeh, F. \&  Fatuzzo, M. 1998, ApJ, 508, 676

\bibitem[]{}
     {Morris, M. \& Serabyn, E.} 1996,
     {ARA\&A} 34, 645

\bibitem[]{}
Martin, C. L. et al. ApJS, 150, 239

\bibitem[]{}
Nayakshin, S. 2006, JPhCS, 54, 208

\bibitem[]{}
Nayakshin, S.,  Cuadra, J. \& Springel, V. 2007, MNRAS, 379, 21

\bibitem[]{}
Nord, M. E. et al. 2004, AJ, 128, 1646

\bibitem[Odenwald \& Fazio(1984)]{1984ApJ...283..601O} Odenwald, S.~F. \&
Fazio, G.~G.\ 1984, \apj, 283, 601


\bibitem[]{}
{Oka, T. et al.  } 2005,
     {ApJ} 632, 882

\bibitem[]{}
{Paumard, T. et al.} 2006,
     {ApJ} 643, 1011

\bibitem[]{}
     {Pierce-Price, D. et al.} 2000,
     {ApJ} 545, L121

\bibitem[]{}
Reid, M. J.  \& Brunthaler, A.  2004, ApJ, 616, 872

\bibitem[]{}
Sanders, R. H. 1998, MNRAS, 294, 35

\bibitem[]{}
Serabyn, E. \& G\"usten, R. 1987, A\&A, 184, 133


\bibitem[]{}
Serabyn, E.  Lacy, J. M.  \& Achtermann, J. E. 1992, ApJ, 395, 166

\bibitem[]{}
Shu, F., Adams, F. C. \& Lizano, S. 1987,  AR\&AA, 25, 23

\bibitem[]{}
Stark, A., Gerhard, O., Binney, J. \& Bally, J. 1991, MNRAS, 248, 14


\bibitem[]{}	
	Tan, J. C. \& McKee, C. F. 2002, in ASP Conference Proceedings, eds: P.A. Crowther,  267, 267


\bibitem[]{}
van der Tak, F.F.S. et al. 
2006, A\&A, 454, L99
                                                                                                           



\bibitem[]{}
Wardle, M. \& Yusef-Zadeh, F. 2007, in preparation

\bibitem[]{}
Yusef-Zadeh, F., Hewitt, J. \& Cotton, W. 2004, ApJS, 155, 421

\bibitem[]{}
     {Yusef-Zadeh, F., Muno, M., Wardle, M. \& Lis, D. C.} 2007b,
     {ApJ} 656, 847


\bibitem[]{}
{Yusef-Zadeh, F., Wardle, M. \& Roy, S.} 2007a, ApJ, 665, L123



\bibitem[]{}
Yusef-Zadeh, F. et al.  2001, 
ApJ,  560, 749


\bibitem[]{}
Zhao, Jun-Hui, Desai, K.,  Goss, W. M. \&  Yusef-Zadeh, F. 1993
ApJ, 418, 235

\bibitem[]{}
Zweibel, E. G. \& Josafatsson, K. 1983,
  ApJ, 270, 511

\end{thebibliography}
\end{document}